\documentclass{article}
\usepackage{arxiv}
\usepackage[utf8]{inputenc} 
\usepackage[T1]{fontenc}    

\usepackage{url}         
\usepackage{booktabs}    
\usepackage{amsfonts}      
\usepackage{nicefrac}      
\usepackage{microtype}     
\usepackage{lipsum}
\usepackage{float}
\usepackage{graphicx}
\usepackage{caption}

\usepackage{subcaption}
\usepackage{titlesec}
\usepackage{multirow}
\usepackage{amsmath}
\usepackage{amssymb}
\usepackage{soul,color}
\usepackage{stackengine}
\usepackage{makecell}
\usepackage{tabularx}
\newcommand\xrowht[2][0]{\addstackgap[.5\dimexpr#2\relax]{\vphantom{#1}}}

\usepackage{xr-hyper} 
\usepackage{hyperref}
\hypersetup{
colorlinks=true, 
}

\newcommand{\norm}[1]{\left\lVert#1\right\rVert}

\usepackage[nodisplayskipstretch]{setspace}

\usepackage{xr}

\makeatletter
\newcommand*{\addFileDependency}[1]{%
  \typeout{(#1)}
  \@addtofilelist{#1}
  \IfFileExists{#1}{}{\typeout{No file #1.}}
}
\makeatother

\newcommand*{\myexternaldocument}[1]{
    \externaldocument{#1}
    \addFileDependency{#1.tex}
    \addFileDependency{#1.aux}
}
\myexternaldocument{supp}

\usepackage{booktabs}

\title{High-throughput discovery of novel cubic crystal materials using deep generative neural networks}
\author{
  Yong Zhao\\
  Department of Computer Science and Engineering\\
  University of South Carolina\\
  Columbia, SC 29201 \\
  \And
  Mohammed Al-Fahdi, Ming Hu *\\
  Department of Mechanical Engineering\\
  University of South Carolina\\
  Columbia, SC 29201 \\
   \texttt{hu@sc.edu}
\And
  Edirisuriya MD Siriwardane, Yuqi Song, Alireza Nasiri\\
  Department of Computer Science and Engineering\\
  University of South Carolina\\
  Columbia, SC 29201 \\
    \And
 Jianjun Hu *\\
 Department of Computer Science and Engineering\\
  University of South Carolina\\
  Columbia, SC 29201 \\
  \texttt{jianjunh@cse.sc.edu} \\
}
\titlespacing {\section}
{0pt}{5.5ex plus 1ex minus .2ex}{3.3ex plus .2ex}
\begin{document}
\maketitle
\begin{abstract}

High-throughput screening has become one of the major strategies for the discovery of novel functional materials. However, its effectiveness is severely limited by the lack of quantity and diversity of known materials deposited in the current materials repositories such as ICSD and OQMD. Recent progress in machine learning and especially deep learning have enabled a generative strategy that learns implicit chemical rules for creating chemically valid hypothetical materials with new compositions and structures. However, current materials generative models have difficulty in generating  structurally diverse, chemically valid, and stable materials. Here we propose CubicGAN, a generative adversarial network (GAN) based deep neural network model for large scale generation of novel cubic crystal structures. When trained on 375,749 ternary crystal materials from the OQMD database, we show that our model is able to not only rediscover most of the currently known cubic materials but also generate hypothetical materials of new structure prototypes. A total of 506 such new materials (all of them are either ternary or quarternary) have been verified by DFT based phonon dispersion stability check, several of which have been found to potentially have exceptional functional properties. Considering the importance of cubic materials in wide applications such as solar cells and lithium batteries, our GAN model provides a promising approach to significantly expand the current repository of materials, enabling the discovery of new functional materials via screening. The new crystal structures finally verified by DFT are freely accessible at Carolina Materials Database \url{www.carolinamatdb.org}.

\end{abstract}

\section{Introduction}
Data-driven accelerated design of new materials is emerging as one of the most promising approaches for addressing the challenges in finding next-generation materials. Currently, one of the main strategies for materials discovery is screening existing materials databases \cite{wollmann2011high,sendek2017holistic,kim2018machine,sorkun2020artificial}. However, such approaches are severely limited by the scale and diversity of the existing structures in the repositories, such as ICSD and Materials Project (MP), which have about ~165,000 and ~125,000 materials, respectively, compared to the almost infinite chemical design space. For example, lithium compounds are widely used in electric vehicles and mobile phone batteries, but there are only ~16,000 different lithium compounds in the MP database, which has been almost exhaustively screened for better lithium-ion battery\cite{liu2020tailoring,nolan2018computation}. Large-scale generation of stable hypothetical crystal structures is strongly needed to significantly expand the current materials repositories in both the quantity and compositional and structural diversity to increase the success rate of high-throughput screening of novel functional materials.

The properties of materials are closely linked to their crystal structures. Traditionally, materials scientists discover new materials by either trial-and-error or heuristic random-guess approaches, both of which are notoriously labor-intensive. One example database is Inorganic Crystal Structure Database (ICSD)~\cite{bergerhoff1987crystallographic}, which collects almost all discovered materials since 1913. To date, only around 165,000 experimental structures are reported in ICSD. Considering the number of elements in the periodic table and their possible combinations, the design space of materials would be infinite combinatorially. Hence, better approaches for new materials discovery is needed.

Several working directions are investigated for the generation of new materials~\cite{oganov2012crystal,oganov2019structure,wang2010crystal,glass2006uspex,dan2020generative,long2020ccdcgan,ren2020inverse,noh2019inverse,kim2020generative}. There are mainly three different ways to generate or discover new crystal structures including doping/element substitution\cite{hautier2011data,liu2020tailoring,shen2020charge,song2020computational}, composition generation plus crystal structure prediction\cite{oganov2019structure}, and generative machine learning models\cite{dan2020generative,noh2020machine,ren2020inverse,jung2018machine,noh2019inverse,kim2020generative}. The element substitution approach is the most widely used strategy. But it is subject to the extremely limited known prototype structures in the database compared to the vast chemical design space. The second approach can exploit the recently developed generative models \cite{dan2020generative} to generate a large number of hypothetical materials compositions and then use crystal structure prediction codes to predict their structures. Many global optimization methods have been developed to search the appropriate compositions and structures, including simulated annealing~\cite{wille1986searching}, basin hopping~\cite{wales1997global}, minima hopping~\cite{goedecker2004minima}, genetic and evolutionary algorithms~\cite{wang2010crystal,glass2006uspex}. Those approaches generally guide the searches towards the local minima of free energy to identify the stable or meta-stable structures either by initial configuration space or chemical composition. However, these crystal structure prediction algorithms are usually too computationally expensive due to their reliance on DFT-based formation energy calculation and can thus only handle relatively simple structures. For complex structures, most of the time, these methods fail to find the ground truth structures corresponding to the global minimum formation energy.

One of the most promising approaches for new materials structure creation is  deep generative machine learning models~\cite{dan2020generative,ren2020inverse,noh2019inverse,kim2020generative,nouira2018crystalgan,court20203,long2020ccdcgan,korolev2020machine}. Both variational autoencoder (VAE)~\cite{noh2019inverse,ren2020inverse,court20203,korolev2020machine} and generative adversarial networks (GAN)~\cite{dan2020generative,long2020ccdcgan,kim2020generative,nouira2018crystalgan} have been adapted for inverse design of inorganic materials with different crystal structure representations. A VAE model contains two parts: an encoder and a decoder~\cite{hinton2006reducing,kingma2013auto,doersch2016tutorial}. The encoder part encodes the crystal structure distribution into a latent space, and the decoder reconstructs the material structures from the latent space. After training, new material structures can be generated by sampling in the latent space. Conversely, a GAN generator model consists of two neural networks: a discriminator (critic) and a generator, both of which are trained simultaneously. The discriminator is trained to differentiate real materials from fake ones generated by the generator, while the generator tries to generate fake materials as real as possible to fool the discriminator. The nash equilibrium achieved by the discriminator and the generator helps a GAN learn the  distribution the materials implicitly. In the past few years, several inorganic materials generative models have been proposed. Those works are limited by their chemical family (e.g. special oxides)~\cite{noh2019inverse,kim2020generative,long2020ccdcgan} or formulas generation~\cite{dan2020generative} or hydrides~\cite{nouira2018crystalgan}. Noh et al.~\cite{noh2019inverse} present a framework for learning a continuous vector for vanadium oxides using VAE, which is trained on a 3D image-like representation to attain the continuous materials space. Two sampling strategies are used in the latent space to generate only $V_{x}O_{y}$ materials. Training the VAE model using 3D grid representation is computationally demanding and memory-hungry. In~\cite{kim2020generative}, Kim and Noh et al. trained a composition family-specific GAN model on the Mg-Mn-O system using the atom coordinates as the representation of materials. The crystal GAN model is composed of three modules: a generator, a critic (discriminator), and a classifier. The critic calculates the Wasserstein distance between real and fake materials~\cite{arjovsky2017wasserstein}. The classifier module ensures that the generator generates desired composition and atom numbers in the unit cell. However, this model can only be used to generate structures of the Mg-MN-O system, and the model quality is limited by the small dataset since there are only limited known compounds of this chemical system. CrystalGAN~\cite{nouira2018crystalgan}, proposed by Nouira et al., consists of a cross-domain GAN model, which maps one hydride system into another using CycleGAN schema~\cite{zhu2017unpaired}. All these works focus on generating materials of a special material system. In a most recent work, Ren et al.\cite{ren2020inverse} proposed a new VAE model that directly uses the atom coordinates and unit cell lattice parameters to encode the structures. To constrain the neural network model behavior, their invertible representation encodes the crystallographic information into the descriptors in both real space and reciprocal Fourier space crystal properties. Their model is trained with 24,785 unique ternary  materials and can generate interesting new structures. However, most of their new structures are generated by perturbing the latent vectors of known materials. Large-scale generation of stable crystal structures remains a challenging problem.  Other than generating materials structures, Dan et.al. \cite{dan2020generative} proposed MatGAN to generate millions of novel materials formulas with chemical validity, which expands the candidates for inverse design of new solid materials. 

In this work, we propose a novel deep generative model called \textbf{CubicGAN} to generate cubic materials structures on a large scale. Ternary materials selected from the OQMD~\cite{kirklin2015open} database are chosen as our training set because of its large size of materials and diverse compositions. In our model, material structures are represented by their lattice parameters, atom coordinates, element embedding, and the space group. The conditions of a specific space group and three elements are fed to the generator to generate desired crystal material structures. 
We trained  ternary and a quarternary GAN models to generate novel cubic (ternary and quarternary) crystal structures of the space groups 216,255,221. Materials of these three space groups consist of 78.5\% of all ternary and quarternary cubic materials in OQMD, covering a majority of known cubic materials space.

Our systematic experiments show that our CubicGAN model can recover not only  many of the known cubic structures but also discover many new materials with new composition prototypes with different anonymous formulas (new prototypes). Additional large-scale DFT based validation has led to the discovery of 506 new cubic crystal materials of new prototypes. The detail of the CubicGAN model will be explained in the following sections. Compared to~\cite{noh2019inverse,kim2020generative,nouira2018crystalgan}, our framework can generate a large variety of materials of different chemical systems. The only work that is similar to ours in terms of variety of materials is~\cite{ren2020inverse}, in which Ren et al. use VAE rather than GAN as the generative model trained with train ternary materials in Materials Project~\cite{jain2013commentary} database. However, their model tends to generate new samples by interpolation. The second major difference is a much simpler representation is used in our work without the momentum
space representations.

Our contributions can be summarized as follows:

\begin{itemize}
    \item We propose a novel GAN model to generate large-scale cubic materials conditioning on the elements and a specified spacegroup. In total, we generate 10 million hypothetical ternary and 10 million quaternary crystal structures for downstream analysis.
    
    \item We perform three stage checks on generated materials and extensively match the generated materials against existing databases. The results show that our method can rediscover a majority of cubic materials in the existing databases. In addition, most of the rediscovered materials from MP are confirmed as stable or meta-stable materials in terms of energy-above-hull.
    \item We perform DFT simulations on 108,897 hypothetical materials, of which 33.8\% novel materials are successfully relaxed. By further analysis, we demonstrate that new crystal structure prototypes (with different anonymized formula types) can be found, such as ABC\textsubscript{6}-216, ABC\textsubscript{6}D\textsubscript{6}-216, and AB\textsubscript{8}C\textsubscript{12}-221.
    \item By further stability verification, 506 new-prototype materials have been generated and confirmed to be stable by phonon dispersion calculation.
\end{itemize}

\section{Methods}

In this work, we focus on training generators of ternary and quarternary cubic crystal structures of three space groups (216, 221, 225) to simplify our model design while ensuring coverage of a majority of the cubic design space. We find that in the OQMD dataset with 813,839 materials, 85.8\% of them are ternary or quaternary materials. In addition, out of all the cubic crystals, 97.8\% of them belong to these three space groups, again covering the majority of the known cubic materials space. These three space groups are selected because we find that for the materials of these three space groups, most of their nonequivalent atom fractional coordinates in the CIF files have a multiplicative factor of 0.25 or belong to this set [0, 0.25, 0.5, 0.75].  So, instead of generating cubic structures with arbitrary real-valued atom coordinates, we only aim to train a cubic material generator that only generates structures whose atom positions are sitting at positions with their fractional coordinate values to be from this set +/-{0, 0.25, 0.5, 0.75}. In this case, the special discrete fractional coordinates are much easier to generate accurately by our deep neural networks. This decision has dramatically simplified our generation model, and thus we choose the training data with these two criteria:ternary and quaternary cubic crystal structures of three space groups (216, 221, 225).

\subsection{Dataset}

We collect the training data from OQMD~\cite{saal2013materials,kirklin2015open}, which is an open-source database of experimental and DFT-calculated materials. Totally 813839 entries are retrieved from version 1.3 of OQMD. Entries calculated with local-density approximation (LDA) are also included. Among them, we successfully build 556,839 and 141,100 POSCAR files for ternary and quaternary materials in the OQMD, of which 505,456 and 127,659 structures belong to cubic crystal systems, respectively. After converting the POSCAR files to symmetrized CIF files, 411,646 ternary materials have three unique nonequivalent atom sites, of which 388,680 materials of cubic crystal systems are found; 129,514 quaternary materials have four unique nonequivalent atom sites, of which 127,523 materials belong to the cubic crystal systems. Table~\ref{tab:general-stat-oqmd} shows the  statistics of OQMD materials distributions. We can find that ternary materials of cubic crystal systems are the largest chuck (91\%) out of all ternary materials. Similarly, it is observed that the ternary cubic structures with 3 nonequivalent sites are 94\% out of all ternary materials with 3 nonequivalent sites. For quaternary materials, these two percentages are 90\% and 98\%, respectively. This means that our CubicGAN model can be used to generate hypothetical cubic materials that are the majority type of known material category.

\begin{table}[H]
\center
\caption{Statistics of OQMD ternary and quarternary materials (Total 813,839)}
\label{tab:general-stat-oqmd}
\begin{tabular}{|c|c|c|c|c|}
\hline
& Ternary  &Ternary cubic  &\makecell{Ternary with 3\\ nonequivalent sites} &\makecell{Ternary cubic with 3\\ nonequivalent sites}\\ \hline
Count& 556,839  &505,456  &411,646  &388,680 \\ \hline
Cubic Percentage & \multicolumn{2}{c|}{505456/556839=91\%} & \multicolumn{2}{c|}{388,680/411,646=94\%} \\ \hline
\hline
& Quaternary  &Quaternary cubic  &\makecell{Quaternary with 4\\ nonequivalent sites} &\makecell{Quaternary cubic with 4\\ nonequivalent sites}\\ \hline
Count& 141,100  &127,659  &129,514  &127,523 \\ \hline
Cubic Percentage & \multicolumn{2}{c|}{127659/141100=90\%} & \multicolumn{2}{c|}{127,523/129,514=98\%} \\ \hline
\end{tabular}
\end{table}

Another key criterion for selecting our training samples is that we only pick cubic structures with three nonequivalent atom positions (in CIF files) for training ternary GAN model (for quarternary GAN, the number is 4). Making this choice allows us to use a unified matrix of dimension ($28\times 3$) to represent all ternary cubic materials (for quarternary materials, the dimension is $27\times 3$ where only one space group is used in this work). For a given material, once we have its nonequivalent positions and space group, the full atom positions within the unit cell can be converted to conventional atom positions by symmetry operations. We have identified 411,646 ternary materials with only three nonequivalent positions, of which 388,680 (94\%) materials belong to cubic crystal systems as shown in Table~\ref{tab:general-stat-oqmd}. Out of these 388,680 materials,  22 space groups are found as shown in supplementary Figure~1,. Among them, the space groups that have the most numbers of materials are Fm$\bar{3}$m and F$\bar{4}$3m (the total portion of these two space groups is 97.2\%). Pm$\bar{3}$m is the third one with only 6,462 samples or 1.7\%.  After removing the duplicate cubic materials within MP and ICSD, almost all the materials (375,749 out of 384,215) with the space groups of Fm$\bar{3}$m, F$\bar{4}$3m and Pm$\bar{3}$m follow this criterion. 

Table~\ref{tab:general-stat-dataset} shows the overall statistics of our finalized training and validation datasets. In total, we have selected 375,749 ternary materials from three cubic system space groups from OQMD to form the OQMD-TC3 (T:Ternary, C-Cubic, 3-three space groups) training dataset: Fm$\bar{3}$m, F$\bar{4}$3m, and Pm$\bar{3}$m each having 186,344, 184,162 and 5,243 materials respectively. These materials together correspond to 249,646 unique formulas. With this diversity of formulas, our CubicGAN model can efficiently learn valid combinations of ternary elements. The unique 84 elements in the datasets are utilized to generate random three-element combinations during GAN training. The same steps are applied to quaternary materials in OQMD. As shown in supplementary Figure~1, materials with space group F$\bar{4}$3m occupies 95\% of the quaternary data. So for training the quaternary GAN model, we only choose materials of space group F$\bar{4}$3m. 

We will use the ternary data from the Materials Project and ICSD databases as validation sets to check the rediscovery rates for our proposed method. We first process the ternary materials in Materials Project database~\cite{jain2013commentary} and ICSD~\cite{bergerhoff1987crystallographic} as we do for OQMD samples to create the MP-TC3 and ICSD-TC3 validation datasets.  In total, 6,545 cubic materials are retrieved, of which the numbers of materials with Fm$\bar{3}$m, F$\bar{4}$3m and Pm$\bar{3}$m are 4576, 520, and 1449, respectively and there are 6,431 unique formulas existing in the whole retrieved data. From the ICSD database, 1,875 cubic materials are found to satisfy our seleciton criteria, of which the numbers of materials are 804, 280, and 791 forf space groups Fm$\bar{3}$m, F$\bar{4}$3m and Pm$\bar{3}$m..  For quaternary materials, the OQMD-QC1 training dataset has 121,008 samples. However, only 39 and 8 quaternary materials are found in MP and ICSD that satisfy our two selection criteria (SeeTable~\ref{tab:general-stat-dataset}). We have removed these samples from our training dataset selected from OQMD by removing the crystal structures with a minor difference of cube lengths from the samples in the validation sets).

\begin{table}[H]
\center
\caption{Statistics of our training data and validation datasets from OQMD, MP and ICSD. Here the cubic materials (6,545+1,875) from MP/ICSD are used as validation set and are excluded from the training set.}
\label{tab:general-stat-dataset}

\subcaption*{Ternary Materials}
\begin{tabular}{|c|c|c|c|c|c|c|}
\hline\xrowht{6pt}
Dataset& Total& Fm$\bar{3}$m & F$\bar{4}$3m & Pm$\bar{3}$m & Unique formula & Unique element \\ \hline  
Training:OQMD-TC3&375,749 & 186,344 & 184,162 & 5,243 & 249,646 &84 \\ \hline
Validation:MP-TC3&6,545 & 4,576 & 520 & 1,449 & 6,431 &84 \\ \hline
Validation:ICSD-TC3&1,875 & 804 & 280 & 791 & 1,034 &84 \\ \hline

\end{tabular}

\vspace{10pt}

\subcaption*{Quaternary Materials}
\begin{tabular}{|c|c|c|c|}
\hline
Dataset& Total& Unique formula & Unique element \\ \hline 
Training:OQMD-QC1&121,008 &39,767  & 56\\ \hline
Validation:MP-QC1&39 &39  & 39\\ \hline
Validation:ICSD-QC1&8 &7  & 12\\ \hline
\end{tabular}

\end{table}

In terms of prototypes in the validation datasets MP-TC3 and ICSD-TC3, supplementary  Table 3 shows details of the existing prototypes for materials that satisfy our selection criteria. We take the prototype "ABC2-225" as an example. Here ABC2 and 225 are the crystal prototype anonymous formula and the space group number used to denote a prototype, and we will use this format in the following content. Overall, the three databases have the same set of prototypes; other than that, MP has an extra one: AB6C6-225. However, only one material (mp-1147668) is found under AB6C6-225 and is unstable. For quaternary materials in OQMD, there are only two prototypes, including ABCD-216, with 121,006 materials and ABCD\textsubscript{6}-216 with two materials. Moreover, we find that quaternary cubic materials distribution is highly biased with 121,018 belonging to space group 216, and only 5674 belonging to space group 225, and no samples found for space group 221. For simplicity, we train the quaternary CubicGAN using only the samples from space group 216 and it then can only generate samples of this space group.

\subsection{CubicGAN Framework}
\begin{figure}[H] 
	\centering
	\includegraphics[width=\linewidth]{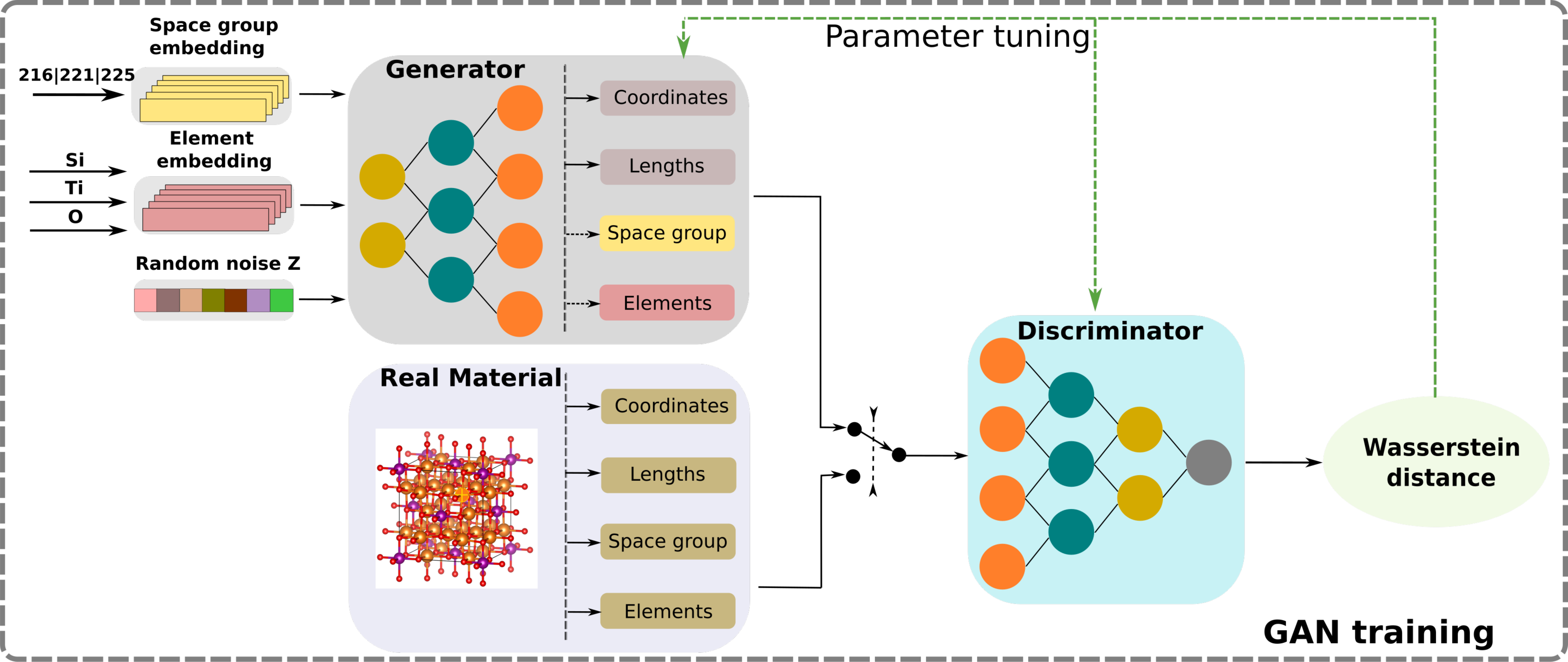}	
  	\caption {The workflow of our CubicGAN framework}
  	\label{fig:framework}
\end{figure}

Figure~\ref{fig:framework} illustrates the main framework of our method. The framework primarily contains two steps: GAN training and material generation. Our goal is to train a generator that learns the distribution from known materials data and then sample from it. To achieve this, the generator is trained to create fake material structures, conditioned on a given space group and a specification of three elements. The three elements are randomly chosen from 84 elements in the dataset. The 84 elements are one-hot encoded and are converted to a $3\times 23$ element matrix by the embedding layer. The parameters of the embedding layer are initialized by 23 element properties as shown in supplementary Table 1. Taking  a randomly selected space group (one-hot encoded), 3-element combinations (one-hot encoded), and random noise $Z$ as inputs, the generator then generates material structures with the specified space group and element constituents. Space groups and elements are mapped into dense vectors by their corresponding embedding layers. The number of atoms for each element does not need to be specified as it can be determined by the space group symmetry operations. The random selections of space groups are based on the portions of three cubic space groups considered in our model: Fm$\bar{3}$m, F$\bar{4}$3m and Pm$\bar{3}$m. The detailed architecture is shown in Supplementary Figure 2.

An input to the discriminator has four parts: nonequivalent atomic coordinates, element properties, unit cell parameters, and space groups as shown in Figure~\ref{fig:framework}. The coordinates part includes the fractional coordinates of three nonequivalent atoms. For three unique elements in each material, each element is represented by 23 properties as shown in the supplementary Table~1. Since the lattice lengths $a,b,c$ are the same in cubic crystals, we only need to use one value to represent it. Three cubic space groups are one-hot encoded. As shown in supplementary Figure 2, four parts are concatenated together to form a tensor with the dimension of $3\times 28$. The input is then forwarded to four 1D convolutional layers, of which the kernel size is $1\times 1$, which is used to capture the implicit relationships among the four parts. We use two CNN layers to reduce the dimension from three to one. Then, a few fully connected layers are used to map them to Wasserstein estimation~\cite{arjovsky2017wasserstein}. The detailed network settings are shown in supplementary Figure~2. In standard conditional GAN, the input of the generator includes the random noise and a condition vector~\cite{mirza2014conditional}.  Here, we add a space group embedding layer and an element embedding layer as shown in Figure\ref{fig:framework} to map the randomly selected one-hot encoded space group (chosen from 216/221/225) and three randomly selected elements (one-hot encoded) into the latent vectors. The reasons for this design are as follows: 1) As only three dominant cubic space groups are used in this work, the combination of atom positions with corresponding elements, unit cell lengths, and one-hot encoded space group symmetry is sufficient to describe a material structure; 2) Using element properties as part of the representation makes the generator learn to generate chemically valid materials, e.g., structures that do not violate Pauling's rules. As our previous work \cite{dan2020generative} shows, the composition constraints can be learned from the compositions of existing materials. Here, our CubicGAN is also configured to learn both implicit compositional as well as structural constraints to help the generator generate only valid ternary or quaternary formulas as much as possible; 3) Our 2D representations of the cubic structures also matches well with the convolutional layers used in the discriminator, in which the convolutional operations can extract implicit relationships among four parts of information.

The generator and the discriminator of the CubicGAN model are trained with the loss function of Wasserstein distance~\cite{arjovsky2017wasserstein} which measures the dissimilarity between distribution differences of real and fake materials. Compared to loss functions used in traditional GAN~\cite{goodfellow2014generative}, Wasserstein distance improves the model stability and prevents the mode collapse. We use the gradient penalty to clip weights in order to improve the stability of training as done by Gulrajani et al.~\cite{gulrajani2017improved}. The penalty of gradient norm with respect to the inputs works as a regularization term to stabilize the training process of the GAN. More formally, our cost function for GAN training is as follows:

\begin{equation}
    L = \underset{\tilde{\mathbf{x}} \thicksim \mathbb{P}_{g}}{\mathbb{E}}[D(\tilde{\mathbf{x}})] - \underset{\mathbf{x} \thicksim \mathbb{P}_{r}}{\mathbb{E}}[D(\mathbf{x})] + \lambda \underset{\hat{\mathbf{x}} \thicksim \mathbb{P}_{\hat{\mathbf{x}}}}{E}[(\norm{\nabla_{\hat{\mathbf{x}}}D(\hat{\mathbf{x}})}_{2}-1)^{2}]
\end{equation}

where $D$ is the discriminator, $\mathbb{P}_{\hat{\mathbf{x}}}$ is the distribution of interpolated samples between the distribution of real materials  $\mathbb{P}_{r}$, and the distribution of generated materials $\mathbb{P}_{g}$. $\lambda$ is the balancing parameter, which is set to 10 in this work. 

After inspecting the generated structures by the GAN, we find that the generated lattice parameter $a$ is often not good enough, leading to overlapping atom clusters. To address this issue, an additional post-processing step introduced to predict the lattice length $a$ using a composition based machine learning model that we recently developed \cite{li2020mlatticeabc}, which achieves a $R^2$ score of 0.979 for cubic lattice $a$ prediction. 

During training the GAN, real materials are randomly picked in batches. With the fused matrix of generated materials as shown in Figure~\ref{fig:framework}, they are fed to the discriminator in a mixed manner. We set the number of iterations of the discriminator per generator iteration as 5. The GAN model is developed using the open-source libraries of TensorFlow~\cite{abadi2016tensorflow} and Keras~\cite{chollet2015keras}. More details regarding model architecture and hyper-parameter setting can be found in Supplementary Table 2 of the supplementary materials. 

\section{Results and Discussion}

\subsection{Performance evaluation of CubicGAN: Validity, Uniqueness, and Rediscovery rate Analysis}

\begin{figure}[H]
    \centering
    \captionsetup{justification=centering}
    \begin{subfigure}{0.5\textwidth}
        \includegraphics[width=\textwidth]{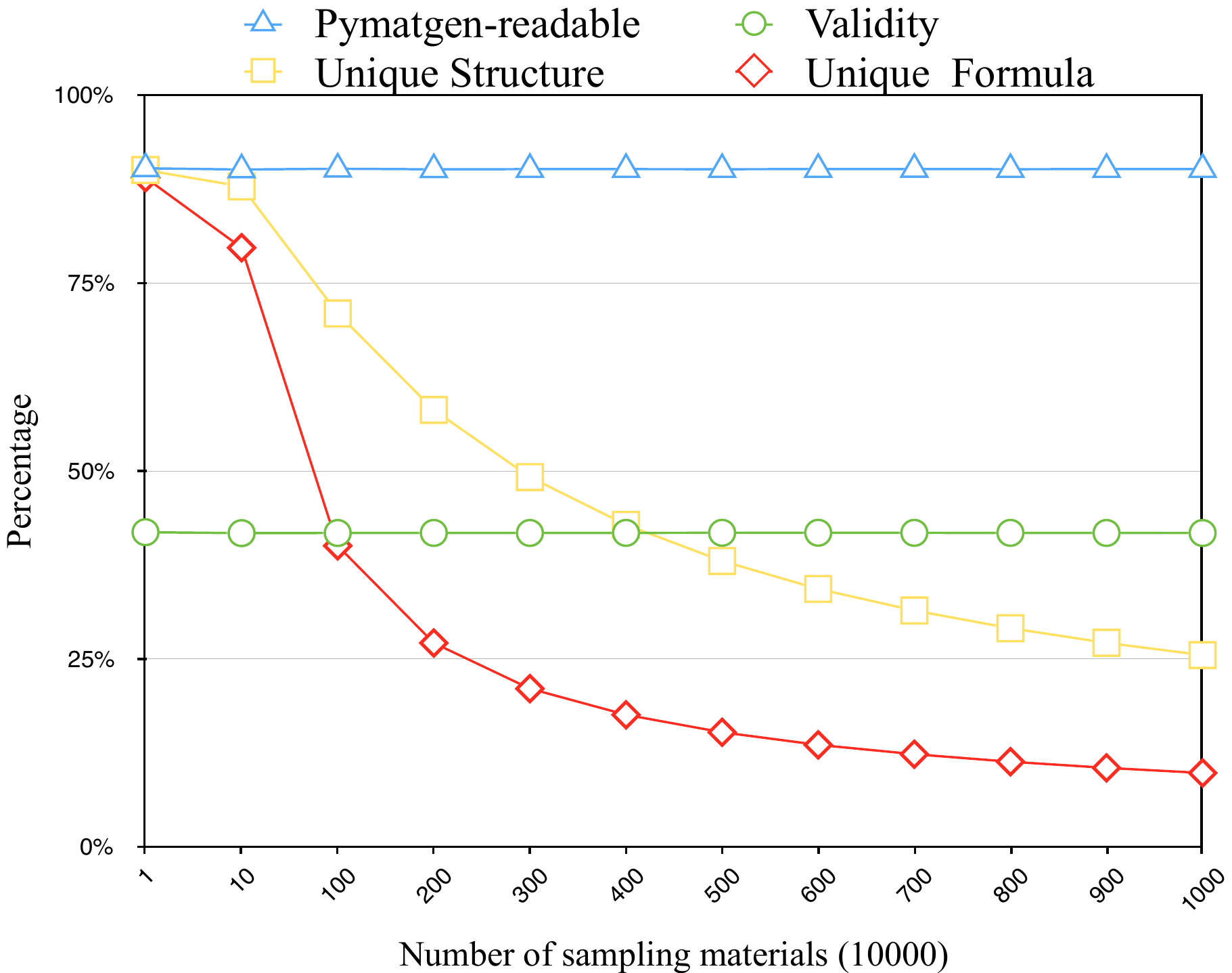}
        \caption{Validity, structural and compositional uniqueness of CubicGAN in terms of number of samplings.}
        \label{fig:gan-trend-mat}
    \end{subfigure}\hfill
    \begin{subfigure}{0.5\textwidth}
        \includegraphics[width=\textwidth]{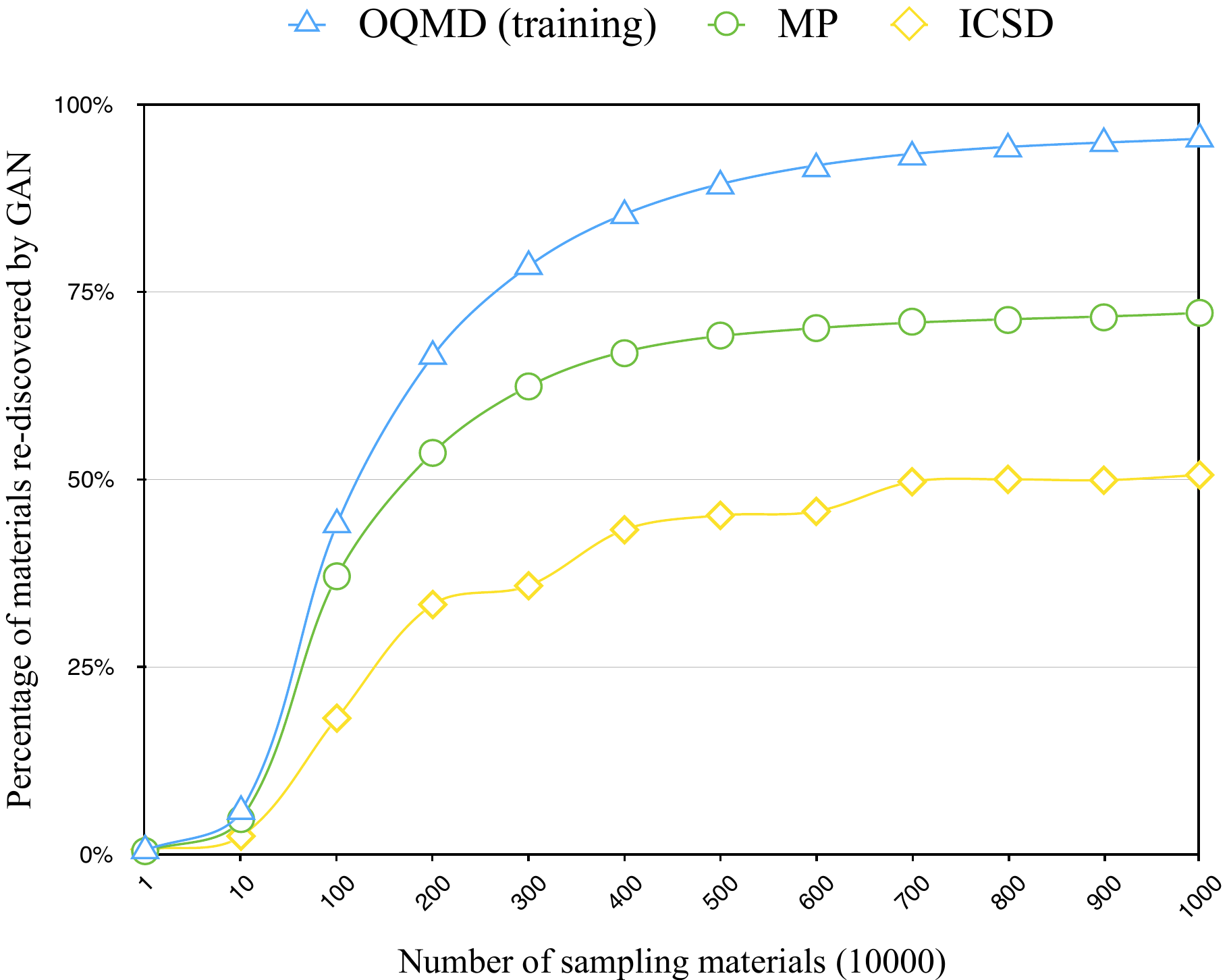}
        \caption{Rediscover rates of CubicGAN in term of the number of sampled materials.}
        \label{fig:trend-overlap}
    \end{subfigure}
    
    \caption{Performance evaluation of CubicGAN}
    \label{fig:perf-eval-gan}
    
\end{figure}

There are three major criteria for evaluating generative models, namely, validity, diversity, and uniqueness~\cite{sanchez2018inverse}. After training the ternary CubicGAN using the OQMD-TC3 dataset, we generate 10 million cubic structures of the specified three cubic space groups (225,216,221). The proportions of the samplings are set as identical to the training set, which is 49.6\%-49.0\%-1.4\% respectively. To evaluate the generation performance, we first check how the percentage of the generated charge-neutral samples changes with respect to the total number of generated samples. The charge neutrality check is based on Pymatgen~\cite{ong2013python} using the common valence values of elements as defined in Pymatgen. As shown in Figure\ref{fig:gan-trend-mat}, the charge-neutral samples' percentage maintains around 41\% over the whole process of generating 10 million samples, which means that when we generate 10 million samples, appropriately we can get 4.1 million charge-neutral samples for downstream screening. 
We then checked how the percentage of the generated samples have pymatgen-readable CIFs (Crystallographic Information File), unique CIFs, and unique formulas, which reflect the diversity and uniqueness of the generator. In Figure~\ref{fig:gan-trend-mat}, the blue line demonstrates the percentage of cifs readable by pymatgen in terms of sampling size, and the sampling size is from ten thousand to ten million. In this work, pymatgen-readable means that CIFs can be recognized as the space group that is assigned to. We can find that the percentage of readable CIF files is stable no matter 
how we run the sampling. After removing the duplicate materials, we calculate the percentage of unique CIFs and unique formulas as denoted by the yellow and red lines in Figure~\ref{fig:gan-trend-mat}. Only those materials that have the same formula and the same corresponding atom positions are considered as duplicates here. It is found that the percentages of the unique CIFs and unique formulas are decreasing and growing flat. From these observations, we believe that our GAN model might have explored the majority of the cubic crystal structure space but have not exhausted it yet.

Another effective way to evaluate the CubicGAN's performance is to check how soon it can rediscover the known cubic crystals in leave-out datasets of existing databases. To do this, in our training dataset, we have removed all the materials of the three cubic space groups (216,225,221) existing in MP and ICSD databases, which are 6,545 and 1,875, respectively. It is interesting to see how many of those leave-out cubic materials can be rediscovered by our GAN model as the sampling size goes from ten thousand to ten million. Figure~\ref{fig:trend-overlap} shows how the rediscovery percentages of the cubic crystals of the three space groups (216,225,221) change as the sampling size increases.

 Figure~\ref{fig:trend-overlap} shows the rediscovery rates over time of sampling. At first, we check how the percentage of the rediscovered cubic samples out of all training samples (blue line) changes while generating more samples. It is found that this training set rediscovery rate increases consistently over the sampling process. It soars quickly to 88\% when the sampling size increases until 5 million samplings are reached. At the end of 10 million samplings, the rediscovery rate reaches 95.5\%. Similar patterns can be observed for the rediscovery rate curve for the MP-TC3 validation dataset, as shown by the green line. With the increasing number of samplings, the rediscovery rate reaches 72.0\%. This saturated percentage is much lower than that of the training set, which is due to MP-TC3 data has different proportions over the three space groups (225,216,221), which are 69.9\%-7.9\%-22.1\% respectively compared to 49.6\%-49.0\%-1.4\% of the training set. Since our generation process is based on the space group proportions of the training set (which focuses on generating candidates of space groups 225 and 216, the 72\% rediscovery rate is close to the percentage of these two types of samples in MP-TC3 (69.9\%+7.9\%=77.8\%). We also find that half of the rediscovered materials in MP-TC3 are stable based on the formation energy and e-above-hull criteria. Details of the stability analysis can be found in  Supplementary Figure 3. The rediscovery rate pattern over ICSD-TC3 is similar to that of MP-TC3 except that the highest rediscovery rate is 50.7\% at the sampling size of ten million, which is close to the percentage of total samples of space groups 225 and 216 (42.9\%+14.9\%=56.8\%). These high rediscovery rates over the training set and the two validation sets demonstrate that our CubicGAN has learned the implicit chemical rules of the cubic structures to generate in a much better way than random sampling. After the sampling size reaches 7 million, the number of materials rediscovered converges, indicating that ten million samplings could be a reasonable size to cover most of the cubic structures since they seem to have almost exhaustively explored the search space of materials that meet our criteria. Therefore, we use ten million samplings for further analysis.

To compare how our CubicGAN performs compared to random sampling or exhaustive enumeration, we calculate the enrichment score for our ternary CubicGAN. As we are searching candidates of three cubic space groups with three unique sites of three distinct elements and the only possible fraction coordinates are 0,0.25,0.5,0.75, the total possibility of configurations are $(4^3)^3*85*84*83*3=466,055,331,840$, which is much larger than the corresponding combinations of the ternary composition space \cite{dan2020generative}. Considering that with 10 million samplings, we have rediscovered 95.5\% of the OQMD-TC3 dataset, the enrichment score is approximately 44,507, which is a significant boosting for generating chemically valid crystal structures compared to exhaustive enumeration.

\subsection{Large scale generation of new cubic crystal materials structures}

\begin{table}[H]
\centering
\caption{Statistics of generated materials}
\label{tab:stat-gen-mat}
\begin{tabular}{|c|c|c|c|}
\hline
 &valid CIFs  &unique formulas  &crystal prototypes  \\ \hline
 Ternary& 2,558,678 &990,319  & 31 (24) \\ \hline
 Quaternary& 5,498,267 & 1,797,592 & 3 (1) \\ \hline
 
 \multicolumn{4}{|c|}{No Lanthanoid and Actinoid}\\ \hline
 
 Ternary& 1,064,650 &403,337  & 31 (24) \\ \hline
 Quaternary& 4,382,130 & 1,431,500 & 3 (1) \\ \hline
\end{tabular}
\end{table}

While rediscovery rate analysis over the MP-TC3 and ICSD-TC3 validation sets have demonstrated the accelerated sampling in cubic structure space, there are only 6,545+1,875=8,420 validation samples plus the 358,840 rediscovered training samples. It is still desirable to check the chemical validity of the remaining 96.33\% generated samples and filter out those promising new materials. With 10 million hypothetical cubic materials, it is impractical to perform DFT calculations for all of them to verify their chemical validity and stability. Here we adopt three stages of validation check to reduce the pool of samples for DFT validation. We use the CGCNN based graph neural network model for formation energy prediction, which was trained with samples from Materials Project database\cite{xie2018crystal}. Then we scan the generated materials in the order of space group match, charge neutrality, and formation energy filtering. The nonequivalent coordinates are transformed by symmetry operations provided by relevant space groups used when generating samples. With the full coordinates set, elements, unit cell parameters (
unit cell length $a$ and angles, which are always 90 degrees in cubic systems), we could write a Crystallographic Information File. The space group check is performed by Pymatgen~\cite{ong2013python} in the first place (we refer to this check as a Pymatgen-recognizable check). If the generated sample cannot be recognized by Pymatgen or the space group analyzed by Pymatgen is not consistent with the space group given to the generated sample, this sample is considered as a failed generated case. As shown in Table~\ref{tab:stat-gen-mat}, in total there are 2,558,678 and 5,498,267 valid ternary and quaternary CIFs have been found from 10 million generations, respectively. From them, candidate materials with charge neutrality and CGCNN-predicted negative formation energy are reserved for further DFT calculations based verification.

A major evaluation of our CubicGAN model is to check whether it can generate new cubic materials with novel prototypes, which are represented by distinct anonymized formulas in Pymatgen. As shown in Table~\ref{tab:stat-gen-mat}, we find that 24 and 1 novel prototypes for ternary and quaternary materials, respectively, have been found in our generated samples that are not existent in our training data. For relieving the burden of DFT calculations, we choose to remove the samples that contain Lanthanoid and Actinoid elements. In total, 1,064,650 ternary materials are left, of which 209,744 materials are of new crystal prototypes. The distribution of prototypes for 1,064,650 materials is shown in Supplementary Figure 4. Similarly, 4,382,130 quaternary materials are left after removing Lanthanoid and Actinoid elements, of which 260,891 materials are of the new crystal prototype (the prototype ABC\textsubscript{6}D\textsubscript{6}-216). Since only two ABCD\textsubscript{6}-216 materials exist in the quaternary training dataset OQMD-QC1, we also include ABCD\textsubscript{6}-216 materials for the downstream DFT analysis considering the huge number of generated ABCD\textsubscript{6}-216 samples (1,655,407). After searching thoroughly in databases of OQMD, MP, and ICSD, only a limited number of materials with ABCD\textsubscript{6}-216 are found, as shown in Supplementary Table 4. Then, we perform charge neutrality check by Pymatgen and CGCNN formation energy filtering on 209,744 ternary materials and 1,916,298 (260,891 + 1,655,407) quaternary materials. While each material might have different atom arrangements in the unit cell that maps to the same space group, in this work, we only choose one of them for DFT calculations. Finally, 17,303 ternary materials and 91,594 quaternary materials are left for DFT optimization. In total, 36847 candidate materials have been relaxed successfully with 14,433 ternary and 22,414 quaternary samples.

\subsection{Discovery of 506 new-prototype stable materials verified with DFT calculations}

After filtering down materials with novel prototypes, we perform DFT optimization on materials with CGCNN-predicted negative formation energy, and we use $\Gamma$ points and mechanic constants to further scale down the successfully relaxed structures. Phonon dispersion is the eventual criterion to determine the stability of structures.

\paragraph{Gamma points and mechanic constants filtering}

The vibrational frequencies at the $\Gamma$~point together with the elastic constants of screened structures were obtained by calculating the Hessian matrix (matrix of the second derivatives of the energy with respect to the atomic positions)~\cite{PhysRevB.65.104104}, which can be done by setting IBRION=6 (NFREE= 4) in VASP run. For cubic structures, the mechanical stability of lattice structures is verified as $C_{11}>0, C_{44}>0, C_{11}>|C_{12}|, C_{11}+2C_{12}>0$, where $C_{ij}$ are components of elastic constant matrix~\cite{ZHANG2017403}. After screening the materials with the mechanical criteria, we further narrow-down the materials by checking the vibrational frequencies at the $\Gamma$ point. All materials with negative $\Gamma$ point frequencies were discarded.

\paragraph{Phonon Dispersion calculation}

After the structures pass the mechanical stability criteria and all $\Gamma$ point frequencies are positive, we further calculate the full phonon dispersions in the first Brinounion zone (BZ). All $2^{nd}$ interatomic force constants (IFCs) of the cubic structures were computed in a 2x2x2 supercell based on their corresponding primitive cell. Then, the phonon dispersions were calculated by using the PHONOPY package~\cite{phonopy} with high symmetry paths $\Gamma \rightarrow X \rightarrow U \rightarrow K \rightarrow \Gamma \rightarrow L \rightarrow W \rightarrow X$ ~\cite{HINUMA2017140}.

\begin{figure}[H]
    \centering
    \captionsetup{justification=centering}
    \begin{subfigure}{0.24\textwidth}
        \includegraphics[width=\textwidth]{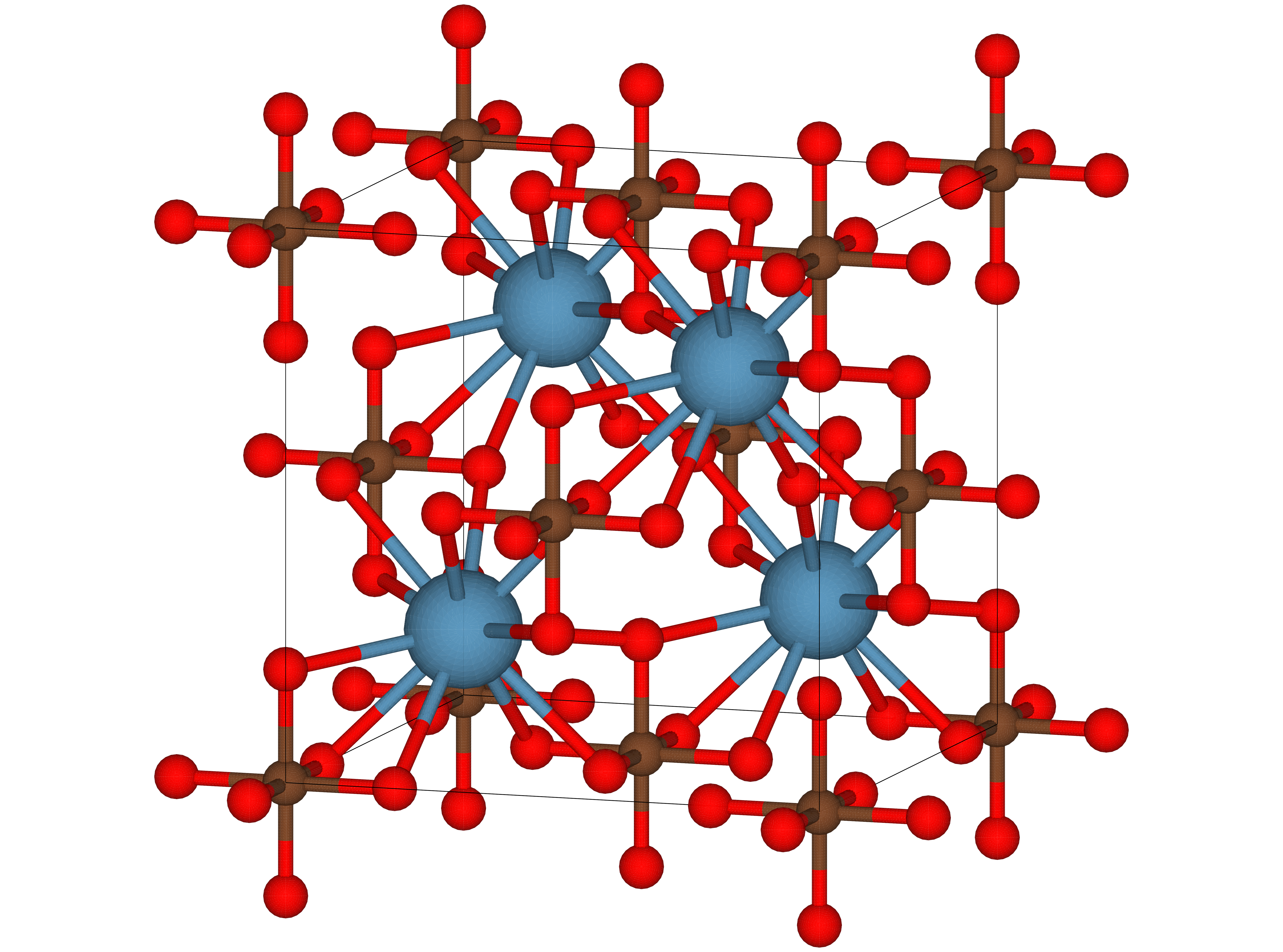}
        \caption*{CaCO\textsubscript{6}}
        \vspace{-3pt} 
        \label{fig:gan-CaCO6}
    \end{subfigure}
    \begin{subfigure}{0.24\textwidth}
        \includegraphics[width=\textwidth]{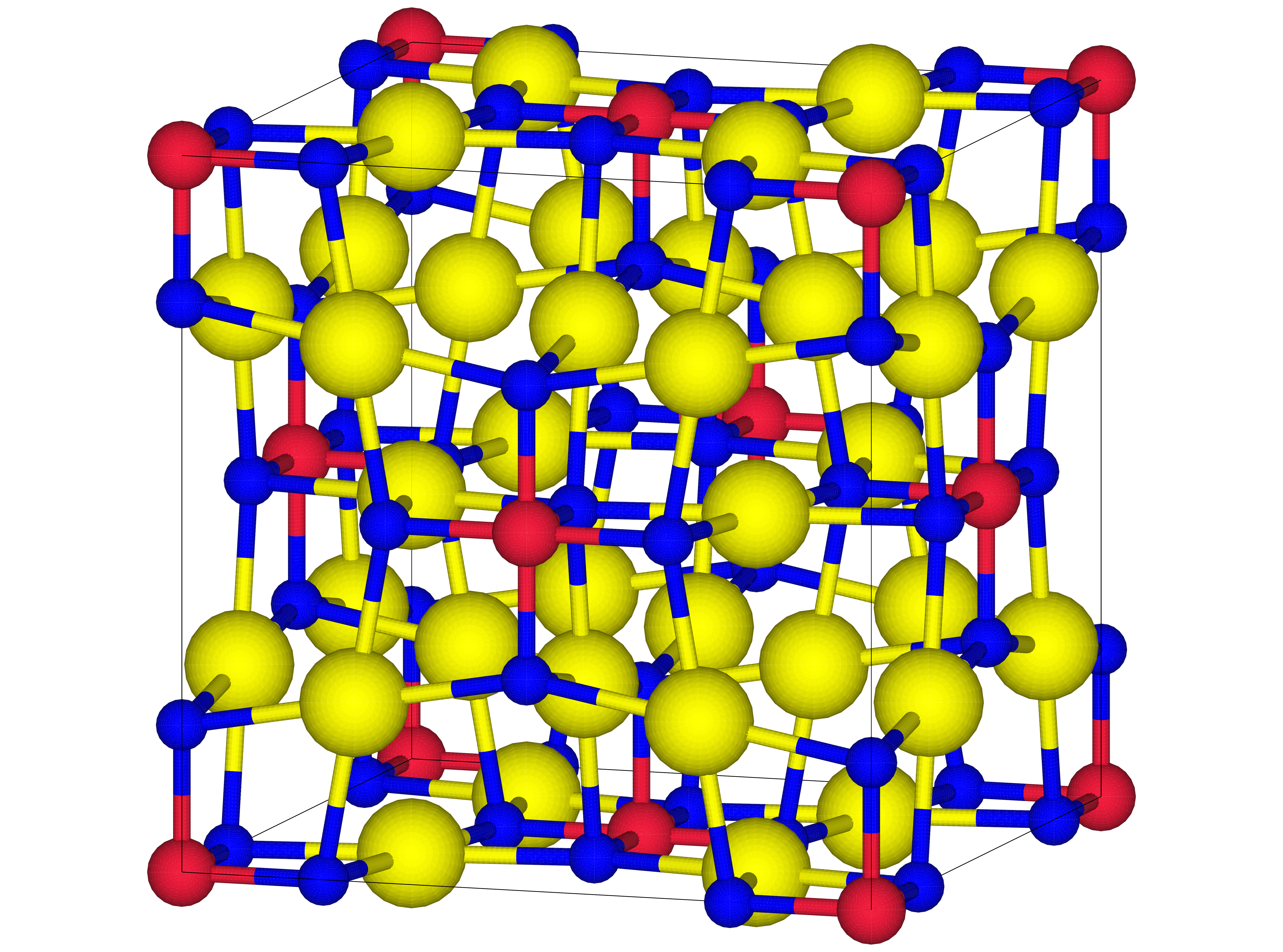}
        \caption*{Li\textsubscript{6}N\textsubscript{6}Cl}
        \vspace{-3pt}
        \label{fig:gan-Li6N6Cl}
    \end{subfigure}
    \begin{subfigure}{0.24\textwidth}
        \includegraphics[width=\textwidth]{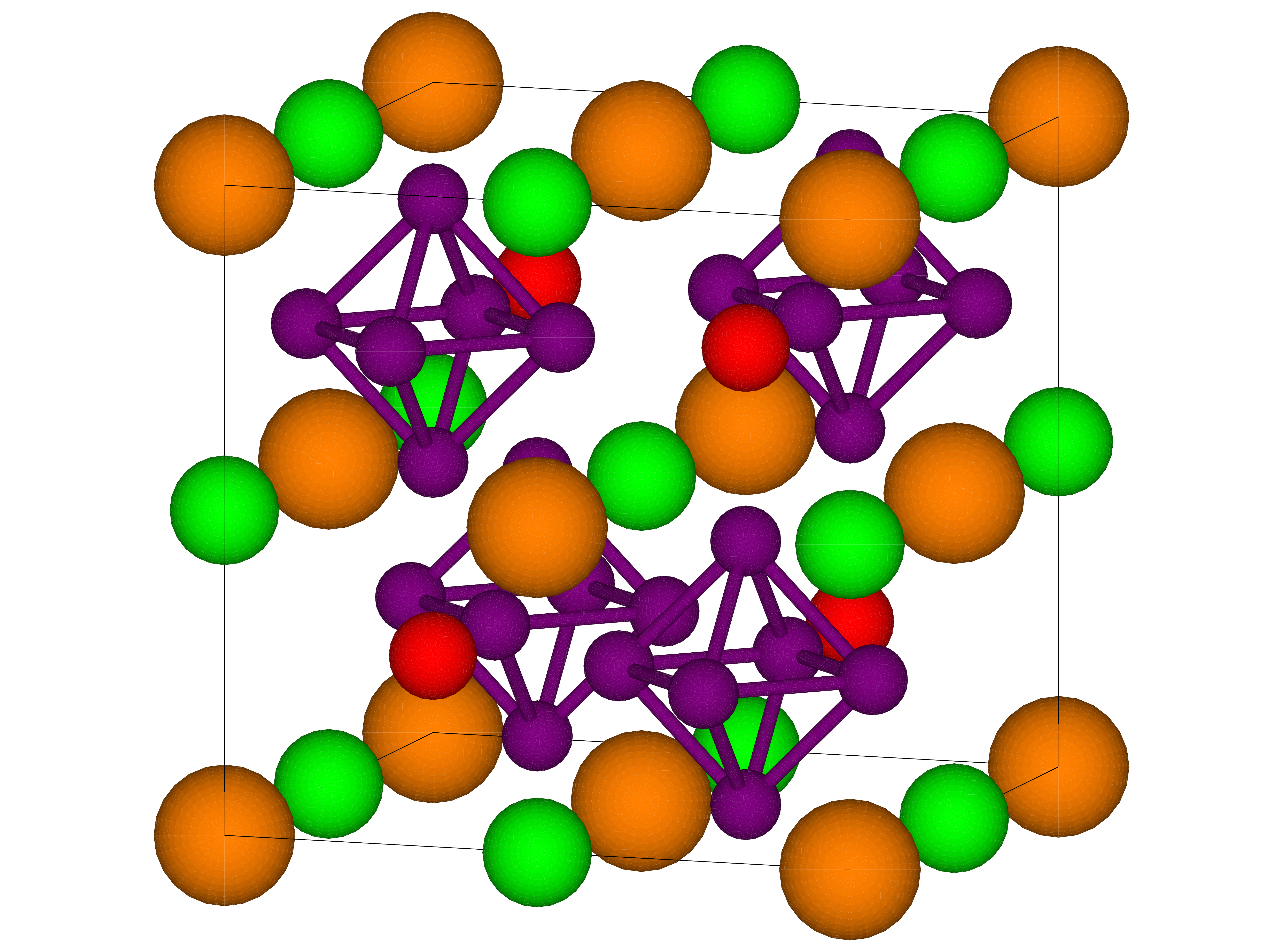}
        \caption*{KYNbSi\textsubscript{6}}
        \vspace{-3pt}
        \label{fig:gan-KYNbSi6}
    \end{subfigure}
    \begin{subfigure}{0.24\textwidth}
\includegraphics[width=\textwidth]{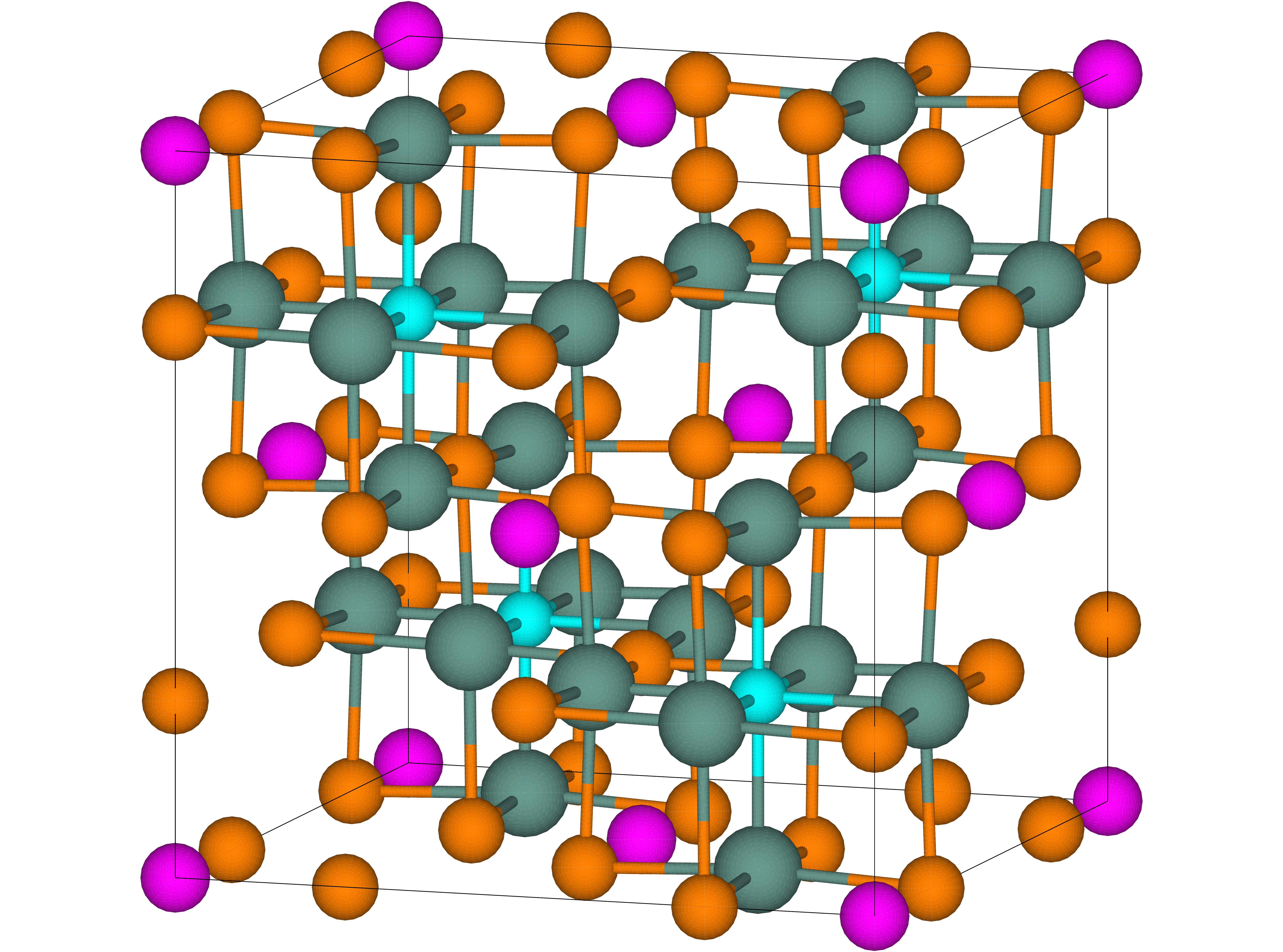}
        \caption*{Y\textsubscript{6}AlTe\textsubscript{6}As}
        \vspace{-3pt}
        \label{fig:gan-Y6AlTe6As}
    \end{subfigure}
    
    \hfill
    \begin{subfigure}{0.24\textwidth}
        \includegraphics[width=\textwidth]{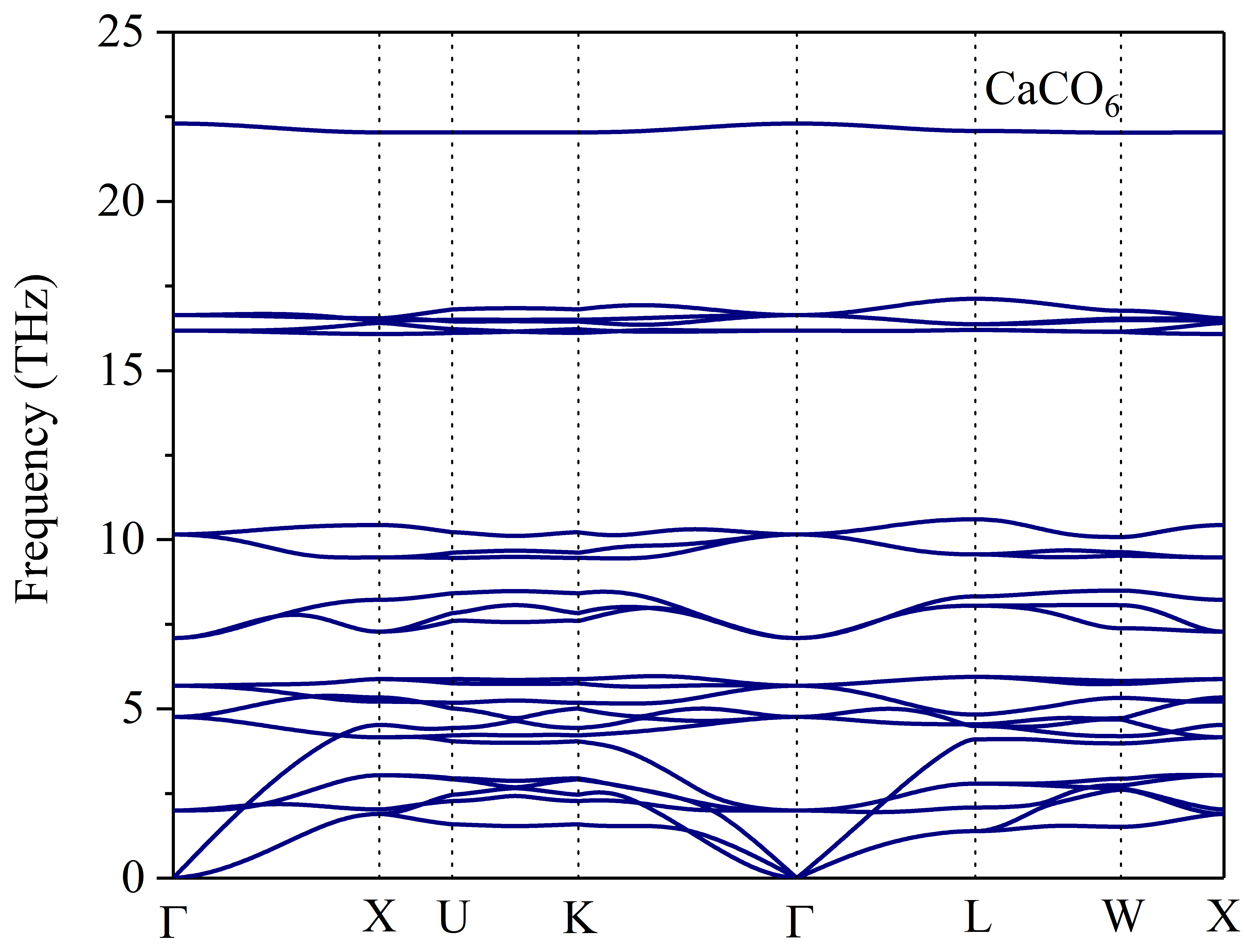}
        \caption{CaCO\textsubscript{6}}
        \vspace{-3pt}
        \label{fig:pd-CaCO6}
    \end{subfigure}
    \hfill
    \begin{subfigure}{0.24\textwidth}
        \includegraphics[width=\textwidth]{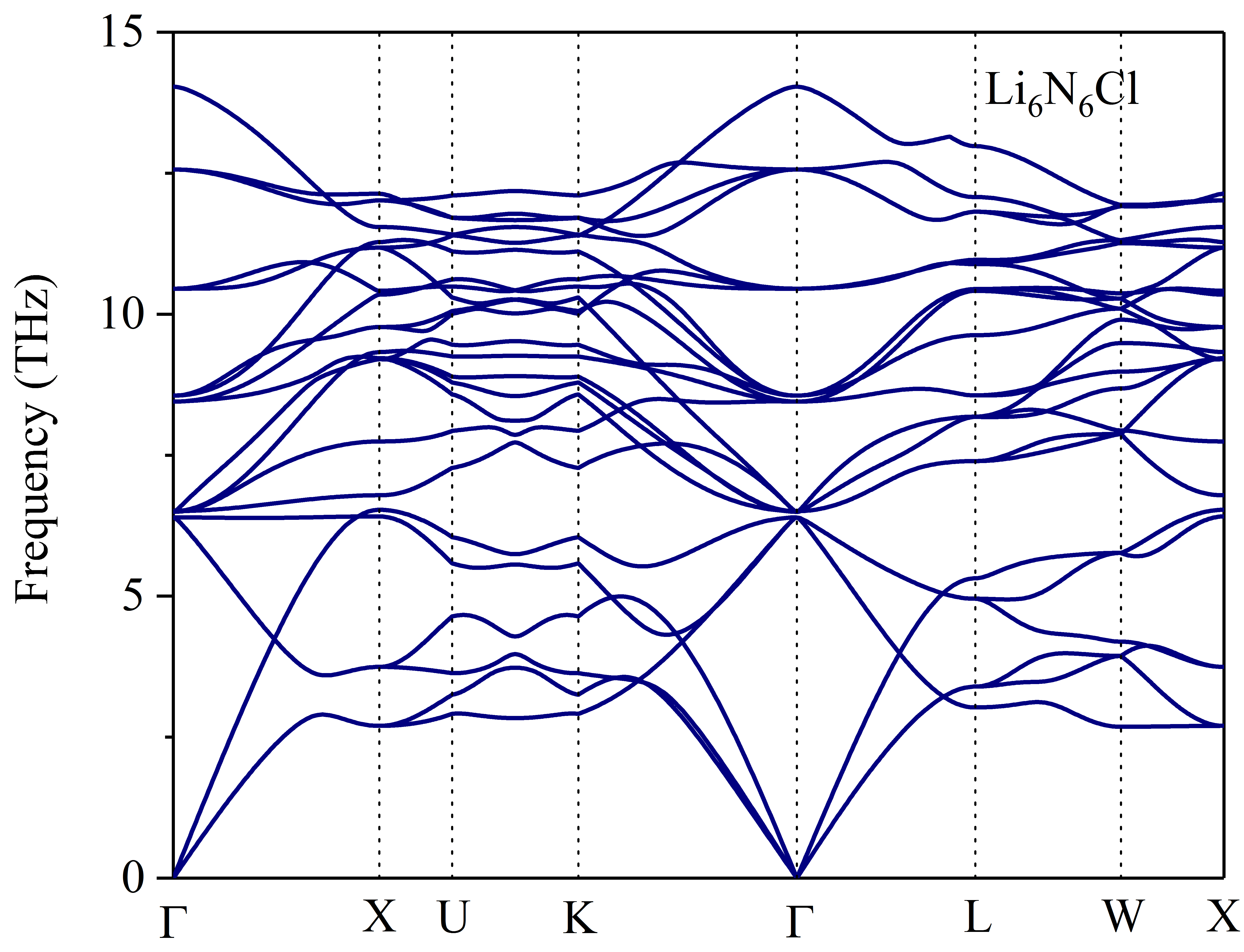}
        \caption{Li\textsubscript{6}N\textsubscript{6}Cl}
        \vspace{-3pt}
        \label{fig:pd-Li6N6Cl}
    \end{subfigure}
    \hfill
    \begin{subfigure}{0.24\textwidth}
        \includegraphics[width=\textwidth]{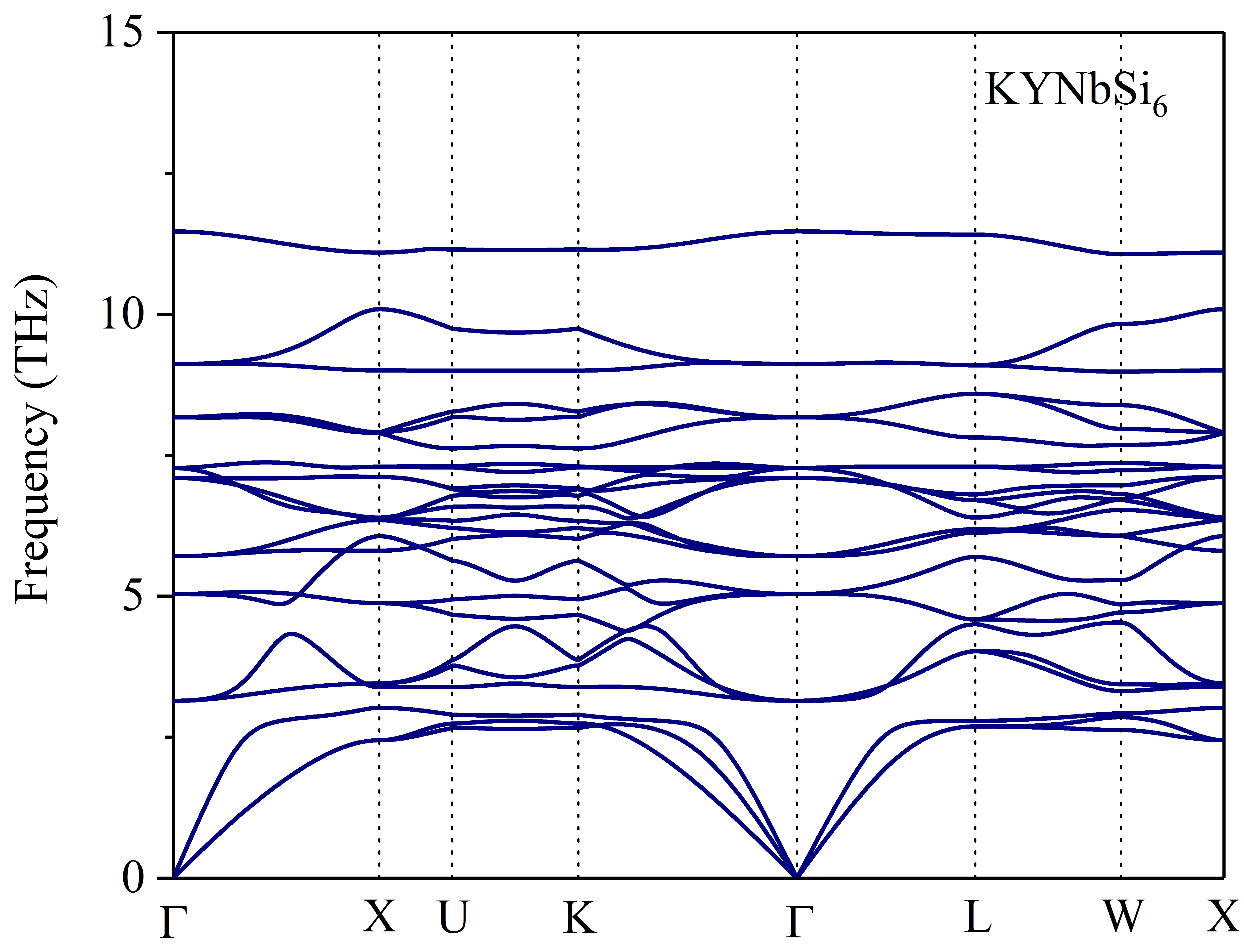}
        \caption{KYNbSi\textsubscript{6}}
        \vspace{-3pt}
        \label{fig:pd-KYNbSi6}
    \end{subfigure}
    \hfill
    \begin{subfigure}{0.24\textwidth}
        \includegraphics[width=\textwidth]{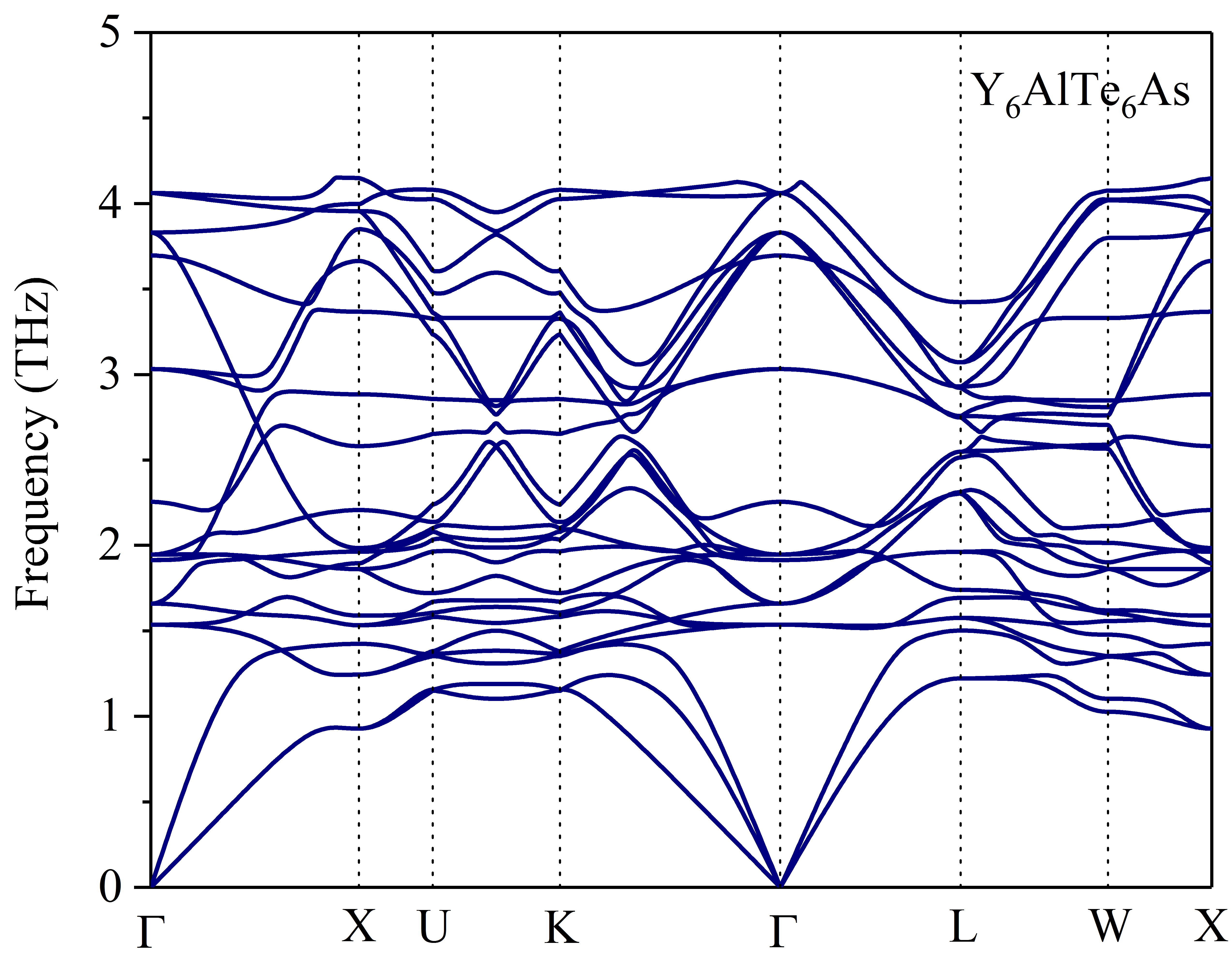}
        \caption{Y\textsubscript{6}AlTe\textsubscript{6}As}
        \vspace{-3pt}
        \label{fig:pd-Y6AlTe6As}
    \end{subfigure}
    
    \caption{Examples of four new prototype materials. Top row is the crystal structures and the second row with corresponding phonon dispersion.}
    \label{fig:exam-mat-novel-prototype}
\end{figure}

In total, four prototypes with stable materials are discovered: ABC\textsubscript{6}-216, AB\textsubscript{6}C\textsubscript{6}-225, ABCD\textsubscript{6}-216, and ABC\textsubscript{6}D\textsubscript{6}-216. The details of the number of materials for each prototype are shown in Supplementary  Figure 5.  To our best of knowledge, ABC\textsubscript{6}-216 and ABC\textsubscript{6}D\textsubscript{6}-216 are novel prototypes that are not in our training dataset, and the validation sets MP-TC3 and ICSD-TC3. Also, the  AB\textsubscript{6}C\textsubscript{6}-225 prototype is not in the training dataset and only one unstable material can be found in MP. However, our method finds 42 stable ones. Two materials of ABCD\textsubscript{6}-216 prototype are in the training dataset, and several others are in MP and ICSD. We expand the datasets by finding 62 stable materials of prototype ABCD\textsubscript{6}-216. Overall, we find 183 stable ternary  materials and 323 stable quaternary materials. Figure~\ref{fig:exam-mat-novel-prototype} shows four newly discovered stable cubic materials with their phonon dispersion curves. The CIF files of the 506 new prototype materials can be found in the supplementary file.

Some interesting features have been observed from the phonon dispersions of newly discovered materials. For instance, a couple of hundred cubic structures we have screened out possess significant but tunable phonon bandgaps (e.g., CaCO\textsubscript{6} as shown in Figure~\ref{fig:exam-mat-novel-prototype}(a)). Such phonon bandgaps could lead to extraordinary hot carrier performance~\cite{chung2014evidence, 9300430, PhysRevB.99.125141, wright2016electron}, which is very promising for their potential application in photovoltaics, nonlinear optics (e.g, ultrashort pulsed lasers), multi-exciton generation devices, and even photocatalysis. Large phonon bandgaps at extremely high frequencies (such as H-containing materials not shown herein) deserve further investigation for their electron-phonon coupling properties~\cite{yue2018electron, yang2018strong, yang2020strong}, which could be beneficial for designing novel superconductors. Also, there are many cubic materials possessing very soft acoustic modes, e.g., the longitudinal acoustic (LA) phonon branch in KYNbSi\textsubscript{6} (Figure~\ref{fig:exam-mat-novel-prototype}(c)), which indicate strong phonon anharmonicity and could be good candidates for waste-heat energy recovery (thermoelectrics). Last but not least, the phonon dispersion of Y\textsubscript{6}AlTe\textsubscript{6} structure exhibits a very large gradient in high-frequency optical phonon modes and thus their phonon group velocities will be very high, which could lead to a significant contribution to the overall thermal transport from these optical modes and thus unusual temperature-dependent lattice thermal conductivity ~\cite{qin2017anomalously}.

\subsection{Visualization of the relationship between new and existing prototypes within the same space group.}

\begin{figure}[H]
    \centering
    \captionsetup{justification=centering}
    \begin{subfigure}{0.5\textwidth}
        \includegraphics[width=\textwidth]{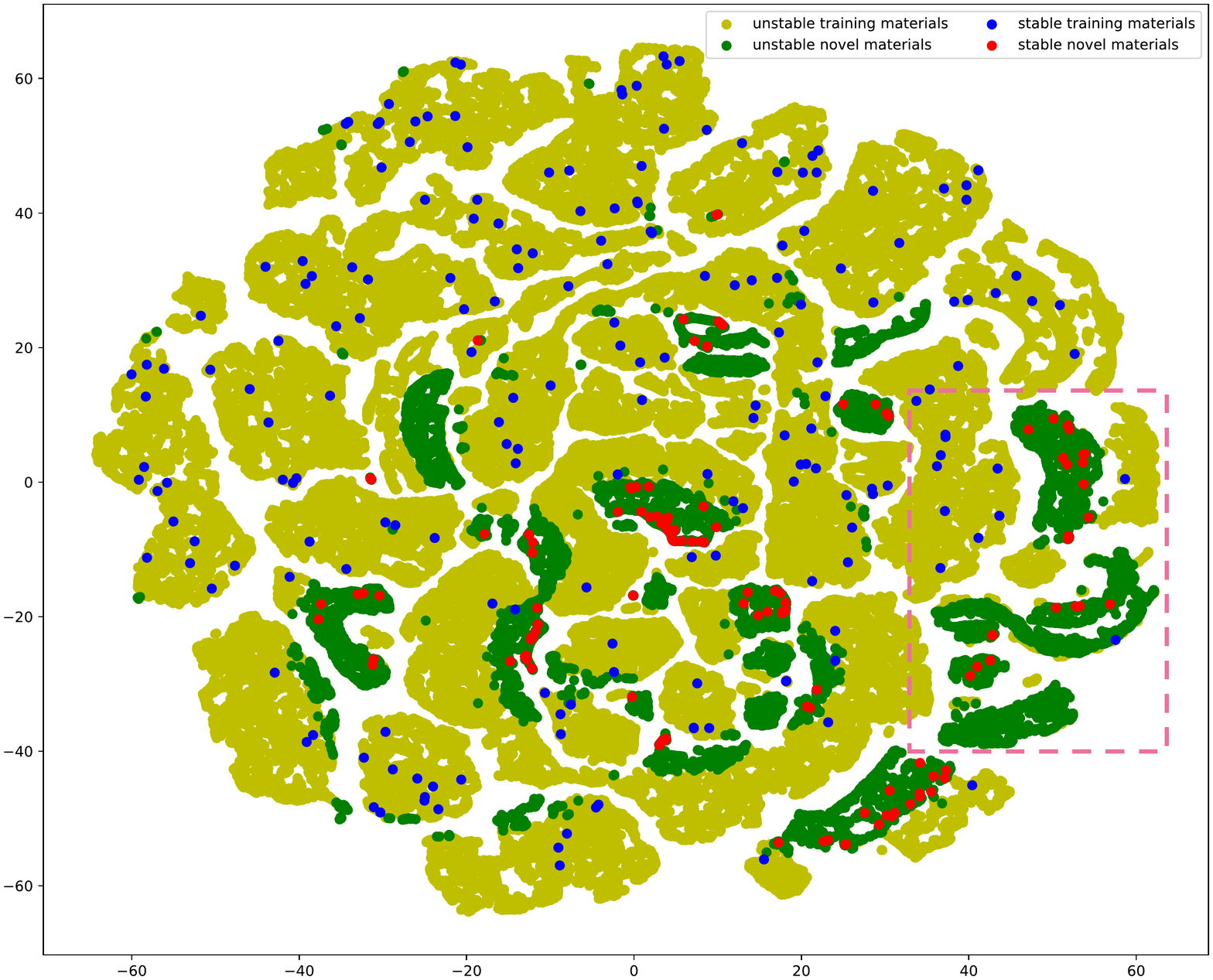}
        \caption{Distribution of ABC\textsubscript{6} materials and the training samples of space group F$\bar{4}$3m}
        \label{fig:abc6-216-cluster}
    \end{subfigure}\hfill
    \begin{subfigure}{0.5\textwidth}
        \includegraphics[width=\textwidth]{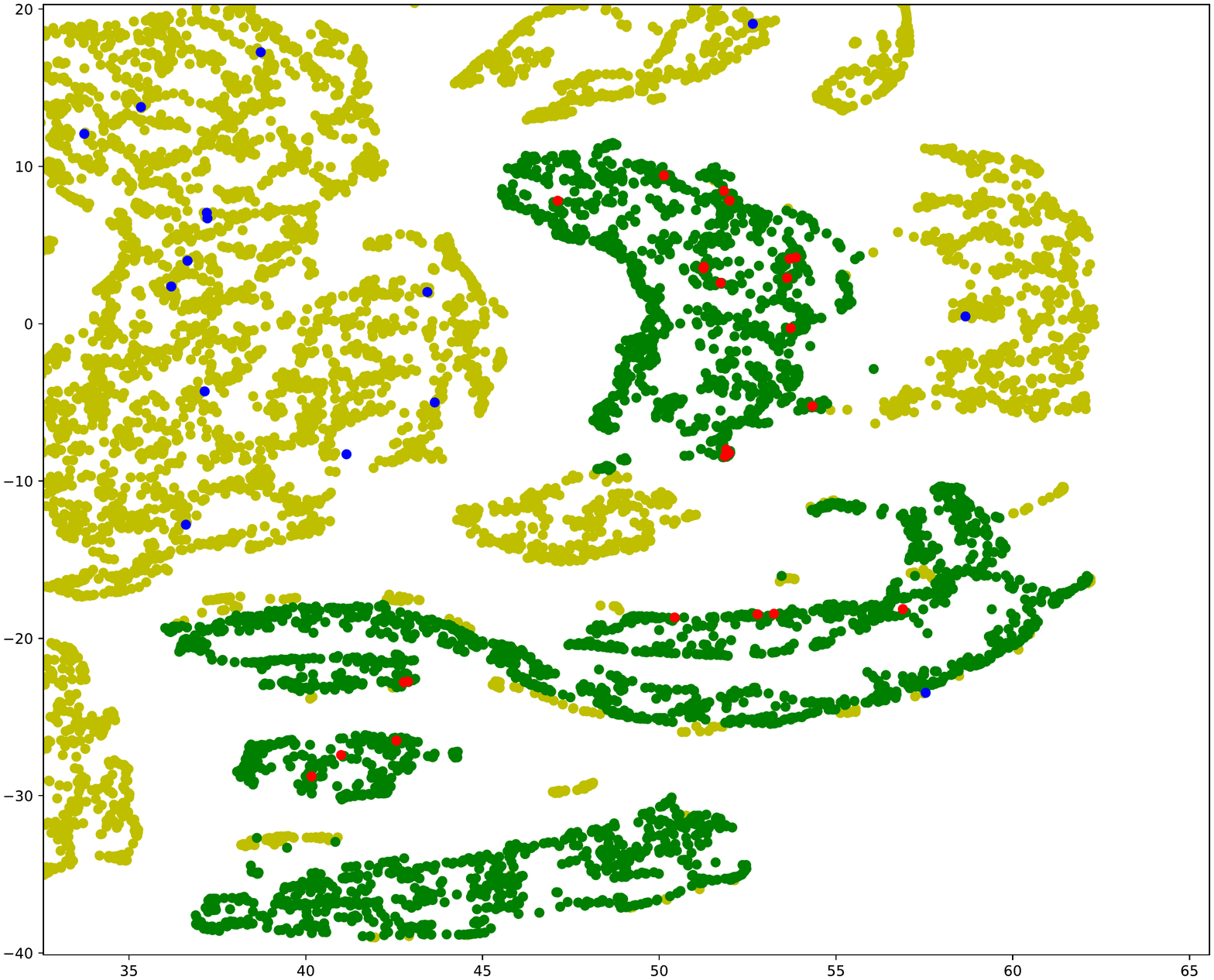}
        
        \caption{Local zoomed region in (a)}
        \label{fig:abc6-216-cluster-local}
    \end{subfigure}\hfill
    
    \caption{Visualization of the structural distributions of the materials in training set and the generated new-prototype ABC\textsubscript{6} materials both belonging to the space group of F$\bar{4}$3m. The new materials form structurally distinct clusters. The 2D space is generated by t-SNE embedding from high-dimensional XRD representations of cubic structures. (a) Overall distributions of existing and new-prototype materials. (b) Zoomed region as marked in sub-figure (a). }
    \label{fig:abc6-new-prototype}
\end{figure}

To qualitatively evaluate how the new-prototype materials are structurally different from existing materials, we represent both sets of materials using simulated (for generated samples) and real 
X-ray diffraction (XRD) spectrum of dimension 901, which is a way to analyze the structures of inorganic materials. The simulated XRDs are calculated using the Pymatgen package ~\cite{ong2013python}.  We then use the t-SNE embedding approach ~\cite{maaten2008visualizing} to map the samples' XRD vectors into 2D space, which are then plotted together to visualize how existing and novel materials of the same space group 216. Figure~\ref{fig:abc6-new-prototype} shows t-SNE embedding of existing materials in the training dataset and newly discovered ABC\textsubscript{6}-216 materials. From Figure~\ref{fig:abc6-new-prototype}(a), we can find that new prototype materials (dark green dots) form distinct clusters, and there are apparent boundaries between known and unknown materials, which indicates that our model can generate materials beyond the scope of existing prototypes with significant structure deviations. Figure~\ref{fig:abc6-new-prototype}(b) shows a zoomed region of  clusters of novel ABC\textsubscript{6}-216 materials, which implies that even samples of the same prototype can form structurally different clusters. For the other three prototypes, the distribution of known structures and our new-prototype structures are shown in Supplementary Figure 6- Figure 9. In Supplementary Figure 6, we find that the new-prototype ( AB\textsubscript{6}C\textsubscript{6}) materials are mostly located at the peripheral regions of know materials clusters, indicating their structural closeness to known structures. In contrast, supplementary Figure 7 shows that materials of two new-prototypes (ABC\textsubscript{6} and AB\textsubscript{6}C\textsubscript{6}) tend to form distinct clusters from known cubic materials in MP-TC3 and ICSD-TC3 validation sets, indicating their structural deviation from known materials. Additionally, for most of these new-prototype clusters, we have identified one or more DFT-verified stable materials. Supplementary Figure 8 shows that materials of new-prototype ABCD\textsubscript{6} form multiple new clusters, each of which contains multiple DFT-verified stable materials. Instead, materials of new-prototype ABC\textsubscript{6}D\textsubscript{6} form much fewer cluster compared to the training set OQMD-TC3.

\section{Conclusion}
Large scale generation of new materials with distinct structures and functions are highly desirable for widely used high-throughput screening based materials discovery. Faced with astronomically large structural design space (compared to the space of the chemical compositions), the generator models have to exploit the implicit sophisticated physicochemical and geometric rules and constraints embedded in the existing crystal materials. Here we propose a novel GAN-based deep generative model for large-scale generation of three major types (space groups:216, 225, 221) of cubic materials structures. Trained with 375,749 ternary cubic crystal structures from OQMD, our CubicGAN model can rediscover most of the known cubic structures as curated over more than 100 years of history within 10 million samplings. Especially, further analysis shows that our GAN model can generate not only new materials of existing prototypes but also new-prototype materials with distinct structural novelty. In total, we have identified 24 new prototypes of cubic materials. With rigorous DFT-based relaxation and phonon dispersion calculation, we have identified and verified 506 new-prototype cubic materials, which are shared via our pubic database. From them, we have already identified several crystal structures with exceptional properties to be exploited in future. Together, our CubicGAN has demonstrated a promising path to large-scale generation and discovery of new materials.

\section{Contribution}

Conceptualization, J.H.; methodology, Y.Z.,J.H.,M.H.; software, Y.Z.,J.H.,Y.S.; validation, Y.Z., M.H., J.H., D.S., M.A.; investigation, J.H., Y.Z., M.H., D.S., Y.S.; resources, J.H.; data curation, Y.Z. and J.H.; writing--original draft preparation, Y.Z., J.H. and M.H.; writing--review and editing, J.H., M.H.,Y.S,A.N.; visualization, Y.Z.,Y.S.; supervision, J.H. and M.H.;  funding acquisition, J.H. and M.H.

\section{Data Availability}
All the training data are downloaded from OQMD and Materials Project website.

\section{Acknowledgement}

Research reported in this work was supported in part by NSF under grant 1940099. This material is based upon work supported by the National Science Foundation Harnessing the Data Revolution Big Idea under Grant No. 1905775. The views, perspective, and content do not necessarily represent the official views of the NSF. We thank Yuxin Li for his help on lattice parameter prediction.

\clearpage

\setcounter{section}{0}

\bibliography{references}
\bibliographystyle{unsrt}

\end{document}